\documentclass[twocolumn,showpacs,aps,pra,amsmath,amssymb,superscriptaddress]{revtex4-2}

\usepackage{graphicx}
\usepackage{dcolumn}
\usepackage{bm,bbm}
\usepackage{hyperref}
\usepackage{subfigure}
\usepackage[ruled,linesnumbered]{algorithm2e}
\usepackage{soul,color,xcolor}   
\usepackage{overpic}

\newcommand{\beq}{\begin{equation}}
	\newcommand{\eeq}{\end{equation}}
\newcommand{\bqa}{\begin{eqnarray}}
	\newcommand{\eqa}{\end{eqnarray}}
\newcommand{\nn}{\nonumber}

\newcommand{\rt}[1]{\sqrt{#1}\,}

\newcommand{\bra}[1]{ \langle{#1} |}
\newcommand{\ket}[1]{ |{#1} \rangle}

\newcommand{\sq}[1]{\left[ {#1} \right]}

\newcommand{\tr}[1]{{\rm Tr}\sq{ {#1} }}

\newcommand{\id}{\mathbbm{1}}

\newtheorem{Theorem}{Theorem}

\newtheorem{Proposition}{Proposition}

\begin{document}
	
	\title{The boundary of Kirkwood-Dirac quasiprobability}
	
		\author{Lijun Liu}
	\email{lljcelia@126.com}
	\affiliation{Department of Mathematics and Computer Science, Shanxi Normal University, Taiyuan 030006, China}
	
		\author{Shuming Cheng}
	\email{drshuming.cheng@gmail.com}
	\affiliation{College of Electronic and Information Engineering, Tongji University, Shanghai, 201804, China}%
	\affiliation{National Key Laboratory of Autonomous Intelligent Unmanned Systems, Shanghai Research Institute for Intelligent Autonomous Systems, Tongji University, Shanghai, 201203, China}


	\begin{abstract}
		
		The Kirkwood-Dirac (KD) quasiprobability describes measurement statistics of joint quantum observables, and has generated great interest as prominent indicators of non-classical features in various quantum information processing tasks. It relaxes the Kolmogorov axioms of probability by allowing for negative and even imaginary probabilities, and thus incorporates the classical probability theory as its inner boundary. In this work, we introduce the postquantum quasiprobability under mild assumptions to provide an outer boundary for KD quasiprobability. Specifically, we present qualitative and quantitative evidence to show that the classical, KD, and postquantum quasiprobabilities form a strict hierarchy, in the sense that joint probability distributions are a strict subset of KD quasiprobability distributions that are a strict subset of postquantum ones. Surprisingly, we are able to derive some nontrivial bounds valid for both classical probability and KD quasiprobability, and even valid for the KD quasiprobability generated by an arbitrary number of measurements. Finally, other interesting bounds are obtained, and their implications are noted. Our work solves the fundamental problems of what and how to bound the KD quasiprobability, and hence provides a deeper understanding of utilizing it in quantum information processing.
		
	\end{abstract}
	

				\maketitle
	
	\section{Introduction} 
	
	Probability theory is a mathematical framework of quantifying uncertainty and modelling random phenomena, which informally assigns real values as probabilities to events, representing their likelihood of occurrence within a defined sample space~\cite{Kolmogorov1956}. It provides the foundation for statistical inference, decision making, and stochastic process, and therefore plays an essential role in quantum mechanics, especially in modelling the measurement process which is intrinsically random~\cite{Nelson2000,Wiseman2009}. However, it is not straightforward to introduce classical probability functions to describe the outcome statistics of simultaneously measuring multiple quantum observables, mainly because measuring one observable typically disturbs the other~\cite{Hall2004,Busch2014,Coles2017}.
	
	Analogy to joint probability distributions for two observables measured in classical mechanics, the Kirkwood-Dirac (KD) quasiprobability~\cite{Kirkwood1933,Dirac1945,Shukur2024} has been introduced to describe measurement statistics of joint quantum observables. It allows for negative and even imaginary probabilities, hence relaxing the standard probability theory formulated by Kolmogorov axioms. Recently, it has found wide applications in the fields of condensed matter physics~\cite{Halpern2017,Halpern2018,Alonso2019}, quantum thermodynamics~\cite{Lostaglio2018,Levy2020,Lostaglio2020,Ghera2024}, postselected metrology~\cite{Shukur2020,Jenne2022,Galdstein2022,Shuker2024B,Salvati2024}, and quantum foundations~\cite{Lostaglio2020,Pusey2014,Budiyono2023A,Schmid2024}, and correspondingly, generated great interest both theoretically and experimentally~\cite{Alonso2019,Shukur2021,Bievre2021,Bievre2023,He2024,Lostaglio2023,Gomez2024,Wagner2024}.
	
	Here, we study the fundamental problems of what is the boundary of KD quasiprobability and how to reveal its boundaries. Inspired by the Popescu-Rohlich box for Bell nonlocality~\cite{Popescu1994}, we introduce the postquantum quasiprobability as an outer boundary of KD quasiprobability, whereas the classical probability theory naturally provides an inner boundary. We provide qualitative and quantitative evidence to show that the classical, KD, and postquantum quasiprobabilities form a strict hierarchy, in the sense that joint probability distributions are a strict subset of KD quasiprobability distributions that are a strict subset of postquantum distributions.
	
\begin{figure}[t]
		\centering
		\subfigure{\includegraphics[width=1.0\linewidth,height=0.55\linewidth]{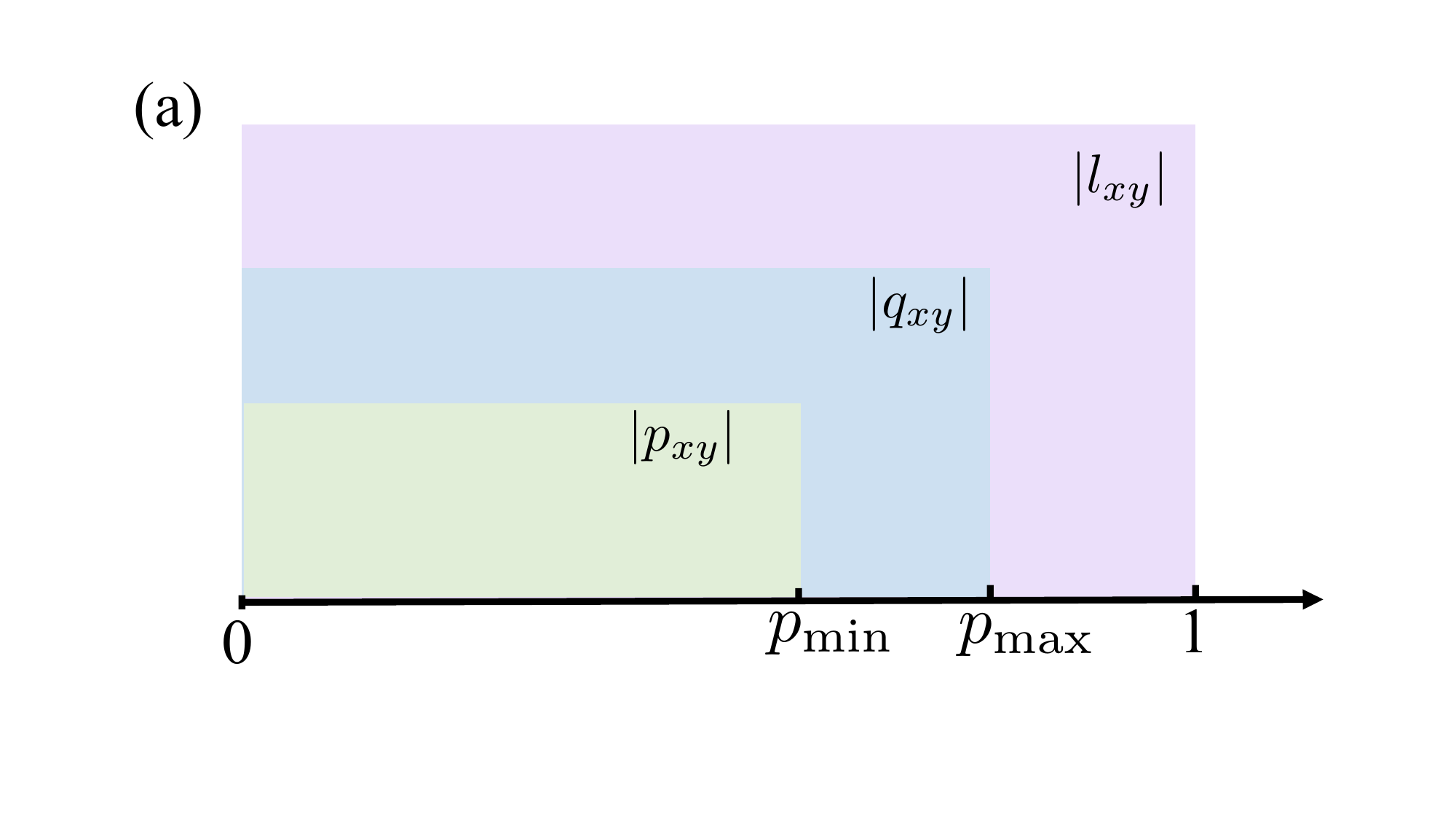}}
		\subfigure{\includegraphics[width=1.0\linewidth,height=0.55\linewidth]{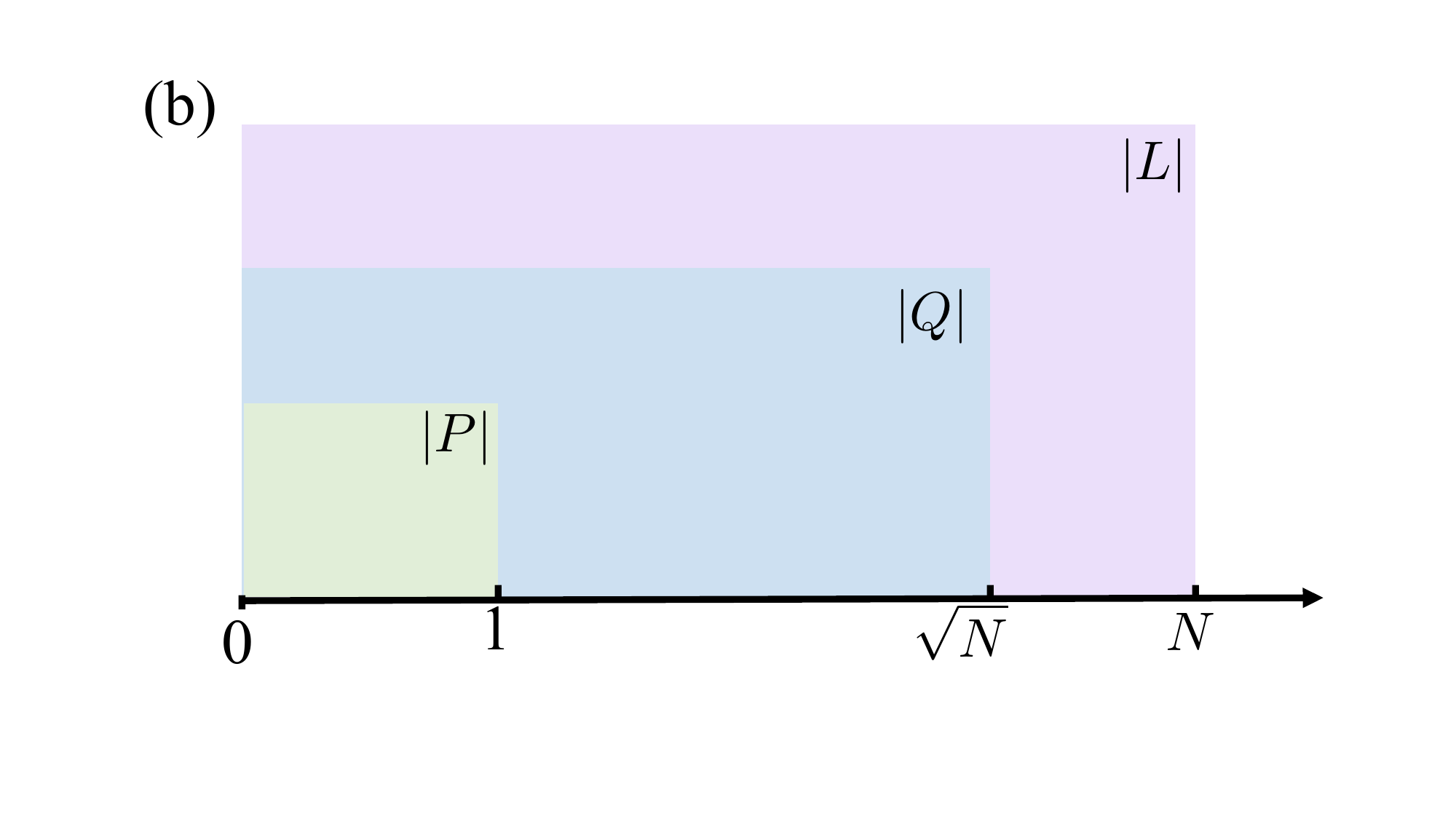}}
	
			\caption{The boundary of KD quasiprobability. (a)  It is shown in Theorem~\ref{Theorem1} that given a pair of marginal probabilities $(p_x, p_y)$ with $p_{\min}=\min(p_x, p_y)$ and $p_{\max}=\max(p_x, p_y)$, the possible region where the global (quasi-)probability lies in is determined by Eq.~(\ref{region}) for classical probability $p_{x y}$, KD quasiprobability $q_{x y}$, and postquantum quasiprobability $l_{x y}$. (b) It is obtained in Theorem~\ref{Theorem2} that the classical, KD, and postquantum bounds are derived as Eq.~(\ref{bounds})
				for any $N$-dimensional (quasi-)probability distribution, with $|P|:=\sum_{x, y} |p_{x y}|$, $|Q|:=\sum_{x, y} |q_{x y}|$, and $|L|:=\sum_{x, y} |l_{x y}|$. These summations could be incomplete by summing over a portion of all possible outcomes.}
		\label{fig:bound}
	\end{figure}

	Particularly, it is first proven in Theorem~\ref{Theorem1} that given a pair of marginal probabilities, denoted by $(p_x, p_y)$ with $p_{\min}=\min(p_x, p_y)$ and $p_{\max}=\max(p_x, p_y)$, the possible region where the global (quasi-)probability lies in is given by
		\beq
	|p_{x y}|\in [0, p_{\min}],~ |q_{xy}| \in [0, p_{\max}], ~|l_{xy}| \in [0, 1], \label{region}
	\eeq
	for classical probability $p_{x y}$, KD quasiprobability $q_{x y}$, and postquantum quasiprobability $l_{x y}$. It is then obtained in Theorem~\ref{Theorem2} that the classical, KD, and postquantum bounds are derived as
	\beq
	\sum_{x, y}|p_{x y}|\leq 1,~ \sum_{x, y}|q_{xy}|\leq \rt{N},~ \sum_{x, y} |l_{xy}|\leq N, \label{bounds}
	\eeq
	for any $N$-dimensional (quasi-)probability distribution, respectively. As displayed in Fig.~\ref{fig:bound}, these results can be utilized as qualitative and quantitative tools to reveal the boundary of KD quasiprobability.

	Importantly, a  state- and measurement-independent bound is further obtained as
	\beq
	\sum_{x, y} |q_{x y}|^\alpha \leq 1, ~~~\alpha\geq 2, \label{square}
	\eeq
	for KD quasiprobability. It is surprising to find that this nontrivial bound still holds for the general KD quasiprobability generated by an arbitrary number of measurements. Finally, the state- and measurement-dependent bounds are derived, and their implications are also noted. All of the results significantly extend the previous study~\cite{Alonso2019,Bievre2021,Bievre2023,He2024,Hall2016} into the regime of general states and measurements.

	This work is structured as follows. In Sec.~\ref{Introduction}, we introduce probability theory formulated by Kolmogorov axioms, KD quasiprobability in quantum theory, and the postquantum quasiprobability proposed by us. In Sec.~\ref{Global}, we present the qualitative evidence to reveal the boundary of KD quasiprobability, by analysing the relation between the global (quasi-)probability and its marginal probabilities. We derive some nontrivial bounds for all KD quasiprobability distributions in Sec.~\ref{universal}, before we construct the quantitative tool to determine the boundary of KD quasiprobability in Sec.~\ref{diffrent}. Finally, our work is concluded with discussions.
	
	\section{Classical probability, KD quasiprobability, and postquantum quasiprobability} \label{Introduction}
	
	We first recap probability theory formulated by Kolmogorov axioms, and then detail KD quasiprobability introduced in quantum theory. Finally, we propose the postquantum quasiprobability that might admit no quantum realization, under the milder assumptions than those of probability and KD quasiprobability.

	\subsection{Kolmogorov axioms of probability}
	
	The mathematical core of probability theory is a triplet $(\Omega, F, P)$, where $\Omega$ is a set of outcomes called as the sample space, $F$ is a set of events called the event space, and probability measure $P$ assigns any event $E$ with some real value. Particularly, the Kolmogorov axioms formulate probability theory with a set of assumptions~\cite{Kolmogorov1956}: \\
	1. {\bf Nonnegativity}: The probability of an event $E$ is a nonnegative real number, i.e., $P(E)\geq 0$ for all $E \in F$. \\
	2. {\bf Unit measure}: The probability of the entire sample space $\Omega$ is 1, i.e.,	$P(\Omega)=1.$ \\
	3. {\bf Additivity}: The probability of any countable sequence of mutually exclusive events $E_1, E_2,\dots,E_n$ satisfies $P(\cup_{i=1}^n E_i)=\sum_{i=1}^n P(E_i).$\\
    These axioms are of fundamental importance to obtain the properties of probabilities. 
	
	Suppose a random variable $X$ with a countable sample space $\Omega$. It follows that for every event $E$ that an outcome $x$ happens, probability measure $P$ is induced as a probability function, with nonnegativity
	 \beq
	 p_x:=P(X=x) \in [0, 1], \label{probability}
	 \eeq
	 and unit sum
	 \beq
	 \sum_{x\in \Omega}p_x=1.
	 \eeq
   Generally, a joint probability function 
   \beq
   p_{xy}:=P(X=x,Y=y) \in [0, 1] \label{classicaljoint}
   \eeq
   can be introduced for two random variables $X$ and $Y$. It admits marginal probabilities functions, in the form of
   \begin{align}
   	p_x=\sum_{y\in\Omega}p_{xy}=P(X=x)\in [0, 1], \label{marginalx} \\
   	p_y=\sum_{x\in\Omega}p_{xy}=P(Y=y)\in [0, 1]. \label{marginaly}
   \end{align}
	Here and elsewhere, these functions are called as probability distributions.

	\subsection{Kirkwood-Dirac quasiprobability}
	
	The Kolmogorov axioms of probability provides the foundation to model the quantum measurement process. For example, consider the projective measurement of quantum observable $O$ on state $\rho$. It is known that each measurement outcome $i$ is obtained with probability
	\beq
	p_i=\tr{\rho\,\Pi_i}\in [0, 1], \nn
	\eeq
	where $\Pi_i=\ket{i}\bra{i}$ denotes the eigenstate projector of $O$. Furthermore, measurement completeness $\sum_i\Pi_i=\id$ ensures $\sum_ip_i=1$, matching well with the probability function as per Eq.~(\ref{probability}).

	 However, it is not straightforward to introduce joint-probability distributions as Eq.~(\ref{classicaljoint}) for two observables measured in quantum theory, mainly due to measurement disturbance that measuring one observable typically disturbs the other. Instead, Kirkwood and Dirac introduced a quasiprobability representation~\cite{Kirkwood1933,Dirac1945}
	\beq
	q_{x y}=\tr{ \rho\, \Pi_x\, \Pi_y},~~~\forall~~x, y\label{KD}
	\eeq
	for two projective measurements $X\equiv\{\Pi_x\}_x$ and $Y\equiv\{\Pi_y\}_y$, where $\Pi_i\Pi_{i^\prime}=\delta_{i{i}^\prime}\Pi_i $ and $\sum_i\Pi_i=\id$ for $i=x, y$. It follows immediately that similar to the classical joint probability~(\ref{classicaljoint}), the KD quasiprobability has unit sum
	 \beq
	\sum_{x, y}q_{x y}=1, \label{unit}
	\eeq
	 and further admits marginal probabilities
	 \begin{align}
	 p_x=\sum_y q_{x y}=\tr{\rho\,\Pi_x}\in [0,1], \label{pa}\\
	 p_y=\sum_x q_{x y}=\tr{\rho\,\Pi_y}\in [0,1].  \label{pb}
	 \end{align} 
 
 	Notably, the nonnegativity assumption in probability theory is relaxed by the above KD quasiprobability~(\ref{KD}) that $q_{xy}$ could be negative and/or nonreal, as product operator $\Pi_x\Pi_y$ is not guaranteed to be Hermitian and positive. More discussions about when and why it is non-positive are referred to Refs.~\cite{Shukur2021,Bievre2021,Bievre2023,He2024,Lostaglio2023, Xu2022,Xu2024,Langrenez2024}. The non-positive KD signals the departure from probability theory, and thus can be regarded as some certain kind of nonclassicality, which has been explored as a source of quantum advantage in various information processing tasks~\cite{Shukur2024}.

 The general measurement can be modelled as positive-operator-valued measure (POVM) $\{E_i\}_i$, where each outcome $i$ is assigned with positive element $E_i\geq 0$ and thus obtained with probability
 $p_i=\tr{\rho\,E_i}.$ As a consequence, it is possible to generalize the KD quasiprobability~(\ref{KD}) to
\beq
q_{x y}=\tr{\rho\, E_x\,E_y}, ~~~\forall~~x, y\label{generalKD}
\eeq
with two POVMs $X\equiv\{E_x\}_x$ and $Y\equiv\{E_y\}_y$. Obviously, it has the same properties as obtained in Eqs.~(\ref{unit})-(\ref{pb}). Again, it is nonclassical, if there exists at least one negative or nonreal $q_{x y}$, or equivalently, the real component of some $q_{x y}$ is strictly smaller than its module, i.e., 
\beq
{\rm Re}(q_{x y})<|q_{x y}|, \label{NS1}
\eeq
which provides a necessary and sufficient criterion to distinguish nonclassical quasiprobabilities from classical ones.

 \subsection{Quasiprobability beyond Kirkwood-Dirac}
 
  The KD quasiprobability relaxes the above probability theory formulated by the Kolmogorov axioms which in turn naturally provides an {\it inner} boundary for KD quasiprobability, in the sense that joint probability distributions are a strict subset of KD quasiprobability distributions. It raises the question of what is the {\it outer} boundary of KD quasiprobability. Indeed, it is subtle to answer this question, as there are so many ways to define quasiprobabilities in quantum theory~\cite{Ferrie2011} and even beyond~\cite{Burgin2012,Xie2015}, by relaxing different assumptions in probability theory.
 
 In this work, we study the above problem under the same assumptions of unit measure and additivity as those of probability theory and KD quasiprobability. Mathematically, we introduce a general form 
 \beq
 l_{x y}:=P(X=x, Y=y),\label{postquantum}
 \eeq
 for two random variables $X$ and $Y$, which is assumed to obey the additivity rule
\beq
\sum_{x, y} l_{x y}=1.
\eeq
Additionally, we explicitly require that its absolute value is always less than 1, i.e.,
\beq
|l_{xy}|=\rt{{\rm Re}(l_{xy})^2+{\rm Im}(l_{xy})^2} \leq 1. \label{module}
\eeq
This assumption is satisfied by joint probability~(\ref{classicaljoint}) and also implicitly assumed in KD quasiprobility~(\ref{generalKD}), to be proven in the following section.

Inspired by the Popescu-Rohlich box for Bell nonlocality~\cite{Popescu1994}, the quantum-realization assumption (e.g., in the form of $\tr{\rho\,E_x\,E_y}$ in Eq.~(\ref{generalKD}) and others in~\cite{Ferrie2011}) is dropped in the general quasiprobability~(\ref{postquantum}), so it is named as postquantum quasiprobability. Similarly, its marginal distributions are obtained as
\beq
p_x:=\sum_y l_{x y},~~~p_y:=\sum_x l_{x y}.
\eeq
Here, they are not required to be probabilities. It is evident that postquantum quasiprobability incorporates KD as a special class, so the classical probability, KD quasiprobability, and postquantum quasiprobability form a hierarchy which is further proven strict in the following sections.

\section{Global quasiprobability v.s. marginal probabilities } \label{Global}

Before proceeding to the qualitative and quantitative evidence that joint probability distributions are a strict subset of KD quasiprobability distributions that are a strict subset of postquantum ones, we first derive a useful relation for KD quasiprobability.

\begin{Proposition} \label{Proposition1}
	The KD quasiprobability~(\ref{generalKD}) satisfies 
	\beq
|q_{x y}|^2\leq  p_x\,p_y,~~~~~\forall~x, y,  \label{KDrelation}
\eeq
where these marginal probabilities are given in Eqs.~(\ref{pa}) and~(\ref{pb}), respectively.
\end{Proposition}

The proof is given as follows. Following first from the Cauchy inequality yields
\begin{align}
	 |q_{x y}|^2=\left|\tr{\rho^{1/2} E_x E_y \rho^{1/2}}\right|^2 \leq \tr{\rho\, E_x^2}\tr{\rho\, E_y^2}. \label{cauchy}
	\end{align}
Note then that $E_x$ and $E_y$ are POVM elements, so $0\leq E_i\leq \id$ and hence $0\leq E_i^2\leq E_i$, which further implies that the expectation values satisfy
\beq
0\leq \tr{\rho\, E_i^2} \leq \tr{\rho\,E_i},~~i=x, y. \label{expectations}
\eeq
Finally, combining Eq.~(\ref{expectations}) with~(\ref{cauchy}) gives rise to the inequality~(\ref{KDrelation}) as desired.
   
It follows from Proposition~\ref{Proposition1} that the module of KD quasiprobability is never larger than 1, i.e., 
\beq
|q_{xy}|\leq 1, \label{moduleq}
\eeq
like the assumption~(\ref{module}) imposed for the postquantum quasiprobability. Moreover, we are able to obtain:

\begin{Theorem}\label{Theorem1}
	The absolute values of the joint probability~(\ref{classicaljoint}), KD quasiprobability~(\ref{generalKD}), and postquantum quasiprobability~(\ref{postquantum}) are bounded by
	\begin{align}
		|p_{x y}| \leq  \min (p_x, p_y), |q_{x y}| \leq  \max (p_x, p_y), |l_{x y}| \leq  1.
	\end{align}
\end{Theorem}
The first inequality follows directly from Eqs.~(\ref{marginalx}) and~(\ref{marginaly}), and the second from Eq.~(\ref{KDrelation}). Further, the possibility that $|q_{x y}| > \min (p_x, p_y)$ is confirmed by choosing $\Pi_x=(\ket{0}+\ket{1})(\bra{0}+\bra{1})/2$ and $\Pi_y=(\rt{3}\ket{0}+\ket{1})(\rt{3}\bra{0}+\bra{1})/4$ on state $\rho=\ket{0}\bra{0}$, and that $ \max (p_x, p_y) < |l_{x y}| =1$ by constructing $l_{00}=l_{11}=1$ and $l_{01}=l_{10}=-1/2$. 

Theorem~\ref{Theorem1} is equivalent to Eq.~(\ref{region}) as summarized in the Introduction, implying that given a pair of marginal probabilities, the classical probability theory, KD-based quantum theory, and postquantum theory allow for different regions where the corresponding global (quasi-)probability lies in. Thus, it gives a qualitative criterion to determine the boundaries among these theories.

\section{Universal bounds for joint probability and KD quasiprobability} \label{universal}

In this section, we obtain some nontrivial bounds valid for both joint probability and KD quasiprobability.These bounds not only have some interest in its own right, and also are useful to derive the quantitative tool to witness the boundary of KD quasiprobability.

\subsection{State- and measurement-independent bounds}

 Summing squared joint probabilities over all possible outcomes yields
\beq
\sum_{x, y} p_{x y}^2 \leq \sum_{x, y} p_x\,p_y=1.\label{upper1}
\eeq
It means that the linear entropy of any probability distribution is smaller than 1. Obviously, the same bounds on the right side of Eq.~(\ref{upper1}) are still valid for all classical KD quasiprobability distributions with $q_{xy}\in [0, 1]$. This immediately implies by contraposition that violating this bound witnesses the KD-nonclassicality.

 Surprisingly, it is found that the bound $1$ can never be violated by KD quasiprobability. Indeed, using the relation~(\ref{KDrelation}) leads to
\begin{align}
\sum_{x,y} |q_{x y}|^2\leq \sum_{x, y} p_x\, p_y=1.   \label{upperl2}
	\end{align}
 It is a state- and measurement-independent bound for KD quasiprobability, significantly strengthening the relation~(\ref{moduleq}). Generally, combining the above bound~(\ref{upperl2}) with~(\ref{moduleq}) yields the nontrivial bounds in Eq.~(\ref{square}) as
 \beq
 \sum_{x, y} |q_{x y}|^\alpha \leq \sum_{x,y} |q_{x y}|^2 \leq 1,~~~\alpha\geq 2. \label{max}
 \eeq
 These bounds admits a similar interpretation that the generalized linear entropy of any KD quasiprobability distribution is always smaller than 1. It is easy to verify that the same bounds can be obtained for all joint probability distributions, signalling no difference between classical probability and KD quasiprobability.
 
 Finally, it follows from the assumption~(\ref{module}) that there exists a trivial bound
 \beq
 \sum_{x,y}|l_{x y}|^\alpha \leq N,~~\alpha\geq 0, \label{trivial}
 \eeq
  for the $N$-dimensional postquantum quasiprobability distribution. Generally, we find that postquantum quasiprobability does not admit such a nontrivial bound like Eq.~(\ref{upperl2}) for KD quasiprobability, as the trivial bound $N=4$ is reached by the case of $l_{00}=l_{01}^*=l_{10}^*=-l_{11}^*=(1+\rt{3}i)/2$ that $|l_{xy}|=1$ and hence $|l_{xy}|^\alpha=1$ for all $x, y=0,1$ and all $\alpha \geq 0$.

 \subsection{State- and measurement-dependent bounds}

 Alternate, a state- and measurement-dependent bound for KD quasiprobability is obtained as
\begin{align}
\sum_{x,y} |q_{x y}|^2&=\sum_{x, y} \left|\tr{E^{1/2}_y\rho E^{1/2}_x E^{1/2}_xE^{1/2}_y}\right|^2 \nn \\
&\leq \sum_{x, y} \tr{\rho\,E_x\rho\,E_y}\tr{E_x\,E_y} \nn \\
&\leq \max_{x, y}\tr{E_x\,E_y} \left( \sum_{x,y} \tr{\rho\,E_x\rho\,E_y}\right) \nn\\
&=\max_{x, y}\tr{E_x\,E_y} \tr{\rho^2}. \label{upper2}
\end{align}
The first inequality follows from the Cauchy inequality, and the second equality from measurement completeness $\sum_x E_x=\id=\sum_y E_y$. As quantum state purity $\tr{\rho^2}$ is no larger than 1, it immediately yields a measurement-dependent bound
\beq
\sum_{x,y} |q_{x y}|^2\leq \max_{x, y}\tr{E_x\,E_y}. \label{measurementdependent}
\eeq

If all POVM elements further satisfy $\tr{E_x\,E_y}\leq 1$, which is obviously satisfied by projective measurements with $\tr{\Pi_x\Pi_y}\leq 1$, then the equation~(\ref{upper2}) gives rise to a state-dependent bound as state purity 
\beq
\sum_{x,y} |q_{x y}|^2\leq \tr{\rho^2}. \label{statedependent}
\eeq
It indicates that the linear entropy of state $\rho$, in terms of two POVMs, is less than its quantum state purity.

It is remarked that the bound~(\ref{upper2}), together with bounds~(\ref{measurementdependent}) and~(\ref{statedependent}), is stronger than the one~(\ref{upperl2}), under the condition $\tr{E_x\,E_y}\leq 1$. Generally, it is not true, by noting the example of two trivial measurements $\id$ and any pure state that $\tr{E_x\, E_y}$ achieves its maximal value $\tr{\id}>1$ and $\tr{\rho^2}=1$. Note also that other interesting bounds are obtained in Refs.~\cite{Budiyono2023B,Budiyono2024A,Budiyono2024B}.

\subsection{Nontrivial bound for KD quasiprobability with multiple measurements} 
 
Mathematically, it is straightforward to generalize the KD quasiprobability~(\ref{generalKD}) with two measurements to 
 \beq
 q_{x_1\,\dots\,x_n}= \tr{\rho\, E_{x_1}\,\dots\,E_{x_n}} \label{multiple}
 \eeq
 with multiple POVMs $X_1=\{E_{x_1}\}, \dots,X_n=\{E_{x_n}\}$. Similarly, it sums over all $x_1,\dots, x_n$ to 1, and admits marginal probability distributions with one single  index $x_i$ with $i=1,\dots,n$. Surprisingly, it is shown in Appendix~\ref{AA} that this generalized KD quasiprobability has the same upper bound obtained in Eq.~(\ref{upperl2}) as
 \beq
\sum_{x_1,\dots,x_n} |q_{x_1\dots x_n}|^2 \leq 1. \label{upper4}
\eeq

If $n$ measurements in the KD quasiprobability~(\ref{multiple}) are grouped into two, denoted by $\{X_1,\dots, X_k\}$ and $\{X_{k+1}, \dots, X_n\}$, then $q_{x_1\dots x_k\,x_{k+1}\dots x_n}$ can be considered as a special type of quantum realization for the postquantum probability $l_{xy}$ defined with $x:=x_1\dots x_k$ and $y:=x_{k+1}\dots x_n$. As a consequence, it is generated by two non-physical measurements on a physical state, with elements $E_x=E_{x_1}\dots E_{x_k}$ and $E_y=E_{x_{k+1}}\dots E_{x_n}$. It is interesting to note that with this form, the postquantum quasiprobability admits a much stronger bound~(\ref{upper4}) than the trivial one~(\ref{trivial}).

\section{Classical, quantum, and postquantum bounds of quasiprobability} \label{diffrent}

The necessary and sufficient condition~(\ref{NS1}) for classical and nonclassical KD quasiprobabilities is equivalent to the witness~\cite{Alonso2019}
\beq
\sum_{x,y} p_{x y}=1=\sum_{x, y}{\rm Re}(q_{x y})<\sum_{x, y}|q_{x y}|. \label{NS2}
\eeq
It provides a quantitative tool, in terms of $l_1$-norm of vector, to determine the boundary between classical and KD quasiprobabilities. We next continue to use this tool to reveal the boundary between KD and postquantum quasiprobabilities. 

For any $N$-dimensional KD quasiprobability distribution, we obtain
\begin{align}
	\left(\sum_{x, y} |q_{x y}|\right)^2 \leq N\left(\sum_{x, y} |q_{x y}|^2\right)\leq N,
\end{align}
by using the Cauchy inequality and Eq.~(\ref{upperl2}). Combining it with Eqs.~(\ref{NS2}) and~(\ref{trivial}) immediately gives rise to the following conclusion.

\begin{Theorem}\label{Theorem2}
	The $N$-dimensional quasiprobability distribution admits the classical, KD, and postquantum bounds, given by
	\beq
\sum_{x, y}	|p_{x y}| \leq 1, ~\sum_{x, y}	|q_{x y}| \leq \rt{N}, ~\sum_{x, y}	|l_{x y}| \leq  N. 
\eeq
\end{Theorem}

Theorem~\ref{Theorem2} provides one quantitative witness to determine the boundaries among the classical, KD, and postquantum quasiprobabilities, in a similar context that the classical, quantum, and non-signalling bounds are derived for the Bell inequality~\cite{Brunner2014}. 

Note that the KD bound in Theorem~\ref{Theorem2} is independent of the specific quasiprobability distribution and hence of the underlying state and measurements, which is further strengthened as follows. Denote $n_{xy}$ by the number of nonzero KD quasiprobabilities $q_{xy}$, so it is easy to obtain $n_{xy}\leq N$ for any $N$-dimensional quasiprobability distribution. Following from Eq.~(\ref{KDrelation}) then yields
\beq
n_{xy}\leq n_x\,n_y \leq N,
\eeq
where $n_i$ refers to the number of nonzero probabilities $p_i$ for $i=x, y$. Finally, combining it with the bound~(\ref{upperl2}) leads to a strengthened bound
\begin{align}
\sum_{x, y}|q_{x y}| \leq \rt{n_{xy}(\sum_{x, y}|q_{x y}|^2)}\leq \rt{n_{xy}} \leq \rt{n_xn_y}
\end{align}
and alternate, using the bound~(\ref{upper2}) yields
\begin{align}
\sum_{x, y}|q_{x y}| &\leq  \rt{n_{xy}\,\tr{\rho^2}\max_{x,y}\tr{E_x\, E_y}} \nn \\
&\leq  \rt{n_x\, n_y\,\tr{\rho^2}\max_{x,y}\tr{E_x\, E_y} }. \label{upper3}
\end{align}
Since the left side is lower bounded by 1 as Eq.~(\ref{NS2}), the support uncertainty relations are obtained as by-product
\beq
n_x\, n_y \geq \max \left(1,~ \frac{1}{\tr{\rho^2}\max_{x, y}\tr{E_x\, E_y}}\right).
\eeq
These results extend the previous study~\cite{Bievre2021,Bievre2023} to the regime of general measurements and general states. 

Finally, it is remarked that the postquantum bound in Theorem~\ref{Theorem2} is a theory-independent bound, as it is derived directly from the assumption~(\ref{module}). This trivial bound is saturated by the example given below Eq.~(\ref{trivial}). Further, if extra assumptions are imposed on the postquantum quasiprobability, then some nontrivial bounds can be obtained. For example, one extra assumption is imposed that the postquantum quasiprobability is real, i.e., $l_{xy}\in [-1, 1]$, and we are able to show in Appendix~\ref{AB} that for any real quasiprobability distribution of dimensionality $N=4$, there exists a strengthened bound $\sum_{x, y} |l_{x y}| \leq 3$, thus signalling the discrepancy between real and complex theories.

\section{Conclusions}

We introduce the postquantum quasiprobability~(\ref{postquantum}) that incorporates the KD quasiprobability~(\ref{generalKD}) as a strict subset that contains the classical probability~(\ref{classicaljoint}) as a strict subset. Then, we present qualitative and quantitative evidence obtained in Theorems~\ref{Theorem1} and~\ref{Theorem2} to reveal the boundaries among the classical, KD, and postquantum quasiprobabilities. We have also obtained some nontrivial bounds such as Eqs.~(\ref{max}) and~(\ref{upper2}) for KD quasiprobability as by-product, which have its own interest.  Our results provide a fundamental insight into the KD quasiprobability and its utility.

There are some interesting problems to be explored in the future work. It is interesting to apply the bounds of KD quasiprobability to study quantum coherence and measurement uncertainty~\cite{Budiyono2023A,Budiyono2023B,Budiyono2024A,Budiyono2024B,Lund2010}. It is also interesting to explore the potential applications in weak value, by utilising close connections between KD quasiprobability and weak value as summarized in~\cite{Shukur2024}. Finally, it is expected to test our nontrivial bounds on the realistic experimental platform.

\section{Acknowledgements}
We greatly thank Dr. Michael Hall for the fruitful discussions and providing the proof in Appendix~\ref{AA}. This work is supported by the Innovation Program for Quantum Science and Technology (No. 2023ZD0301400) and the National Natural Science Foundation of China (No. 12205219).

\appendix

	\section{Proof of the upper bound~(\ref{upper4})}\label{AA}
	
	For the case of two general measurements described by positive-operator valued measure (POVM) $\{E_x\}$ and $\{E_y\}$, we have
	\begin{align}
		\left| \tr{\rho\, E_x\, E_y} \right|^2& = \left| \tr{(\rho^{1/2} E_x\, E_y) \rho^{1/2} } \right|^2 \nn \\
		&\leq  \tr{\rho^{1/2} E_x\,E_y\, E_y\, E_x\rho^{1/2}}\tr{\rho} \nn \\
		&	= \tr{(E_x\, \rho\, E_x) E_y^2 }  \nn \\
		&	= \tr{ \rho'\, E_y^2} \tr{E_x\, \rho\, E_x} \nn\\
		&\leq \tr{\rho' \,E_y} \tr{E_x\, \rho\, E_x},\label{rhop}
	\end{align}      
	where $\rho' := (E_x\, \rho\, E_x) / \tr{E_x\, \rho \,E_x}$. So, summing this inequality over the elements of the POVM $\{E_y\}$ gives
	\begin{align}
	\sum_y 	\left| \tr{\rho\, E_x\, E_y} \right|^2 &\leq  \sum_y \tr{\rho^\prime \,E_y} \tr{E_x\, \rho\, E_x} \nn \\
	&=\tr{\rho\, E_x^2}. \label{sumb}
	\end{align}
	Then, using the fact that $E_x$s are the POVM elements yields $0\leq E_x\leq \id$ and hence $0\leq E^2_x\leq E_x$, which further implies
	\beq
	\sum_{x,y} \left| \tr{\rho\, E_x\, E_y} \right|^2 \leq \tr{\rho} = 1.
	\eeq
	
	One can do the general case of $n$ measurements in essentially the same way, via an inductive process. Specifically, consider general operators $E$ and $P$, where $E$ is a POVM element, and note first that
	\beq
	\tr{ \rho P E^2 P^\dag} = \tr { \rho_P E^2}\tr{P^\dag \rho P}
	\eeq
	for $\rho_P := P^\dag \rho P / (P^\dag \rho P)$, and use the same reasoning as above to obtain
	\beq
	\sum_x \left| \tr{ \rho P E_x^2 P^\dag} \right|^2 \leq \tr{\rho P P^\dag}
	\eeq   
	for any POVM $\{E_x\}$.  Defining $P_n$ to be the product
	\beq
	P_n= E^{(1)}_{x_1} \dots E^{(n)}_{x_n}
	\eeq
	of elements for $n$ POVMs $\{E^{(1)}_{x_1} \}, \dots, \{ E^{(n)}_{x_n} \}$, and applying Eq.~(\ref{sumb}) inductively, then gives
	\begin{align}
		\sum_{x_1,\dots, x_n} \tr{\rho P_n P_n^\dag } \leq& \sum _{x_1,\dots, x_{n-1}} \tr{\rho P_{n-1} P_{n-1}^\dag}\nn \\
		\leq& \dots \leq \tr{\rho}=1. \label{sumn}
	\end{align}
	Finally, substitute 
	\begin{align}
		| \tr{ \rho P_n} |^2  &= \left|\tr{(\rho^{1/2} P_n) \rho^{1/2}} \right|^2
		\leq  \tr{\rho P_n P_n^\dag} \tr{\rho} \nn \\
		&= \tr{\rho P_n P_n^\dag}
	\end{align}
	into the above equation~(\ref{sumn}), to give 
	\beq
	\sum_{x_1,\dots,x_n} |q_{x_1\dots x_n}|^2 \leq 1  \nn 
	\eeq
	as desired. 
	
	\section{The strengthened bounds for the real postquantum quasiprobability} \label{AB}
	
	 It is exemplified in the main text that the following postquantum quasiprobability distribution with
	 \begin{align}
	 	l_{00}&=\frac{1+\rt{3}i}{2},~~~ l_{01}=\frac{1-\rt{3}i}{2}, \nn \\
	 	l_{01}&=\frac{1-\rt{3}i}{2},~~~ 	l_{11}=\frac{-1+\rt{3}i}{2}, \nn 
	 		\end{align}
 		saturates the trivial bound 
	\beq
	\sum_{x,y}|l_{x y}|=\sum_{x, y} 1 = 4.
	\eeq
	If the postquantum quasiprobability is further assumed to be real, i.e., 
	\beq
	l_{00}, l_{01}, l_{10}, l_{11} \in [-1, 1], \label{real}
	\eeq
	then we show that there exists a strengthened bound 
	\beq
	\sum_{x, y} |l_{x y}| \leq 3. \label{strengthen}
	\eeq
	
	The proof is detailed as follows. First, we only need to consider the case where at least one quasiprobability element is negative, because $\sum_{x,y}|l_{x y}|=\sum_{x, y} l_{x y}=1$ for all nonnegative $l_{xy}$. Thus, assume that $l_{00}$ is negative, and then we analyse the bound case by case. \\
	1) If there is a unique negative quasiprobability $l_{00}$, we have
	\begin{align}
	 |l_{00}|+ |l_{01}|+|l_{10}|+|l_{11}|=&-l_{00} +l_{01}+l_{10}+l_{11} \nn \\
		=&1-2\,l_{00}\leq 3.
		\end{align}
	 The second equality follows directly from the unit sum assumption and the final inequality from Eq.~(\ref{real}). This bound is saturated by the distribution $(-1, x, y, 2-x-y)$, with $x, y\in [0, 1]$ and $x+y\geq 1$. \\
	 2) If there are two negative quasiprobabilities, we obtain
	 	\begin{align}
	 	|l_{00}|+ |l_{01}|+|l_{10}|+|l_{11}| =&-l_{00} -l_{01}+l_{10}+l_{11} \nn \\
	 	=&2\,l_{10}+2\,l_{11}-1 \nn \\
	 	\leq& 2+2-1 = 3
	 \end{align}
 for negative $l_{01}$. It is saturated by $(x, -x-1, 1, 1)$, with any $x\in [-1, 0]$. Performing permutation of $l_{01}$ and $l_{10}$ in the above equation yields the same bound for negative $l_{10}$, and similarly, the permutation of $l_{01}$ and $l_{11}$ leads to the same conclusion for negative $l_{11}$. \\
3) We are able to obtain
	\begin{align}
	|l_{00}|+ |l_{01}|+|l_{10}|+|l_{11}|=&-l_{00} -l_{01}-l_{10}+l_{11} \nn \\
	=&2\,l_{11}-1\leq 1<3
\end{align}
for unique nonnegative $l_{11}$, and similarly for either $l_{01}$ or $l_{10}$.

Finally, it is impossible to have four non-positive quasiprobabilities, so we complete the proof of Eq.~(\ref{strengthen}).

\end{document}